\documentstyle[11pt,emulapj,epsfig]{article}

\begin{document}

\title{Magnetic Helix Formation Driven by
Keplerian Disk Rotation in an External Plasma
Pressure ---  The Initial Expansion Stage}

\author{H. Li\altaffilmark{1}, R.V.E. Lovelace\altaffilmark{2},
J.M. Finn\altaffilmark{3} and S. A. Colgate\altaffilmark{4}}

\altaffiltext{1}{Applied Physics Division,  MS B288,
Los Alamos National Laboratory, Los Alamos, NM 87545;
hli@lanl.gov }
\altaffiltext{2}{Department of Astronomy, 
Cornell University, Ithaca, NY 14853; rvl1@cornell.edu}
\altaffiltext{3}{Plasma Theory, Los Alamos
National Laboratory,  Los Alamos, NM
87545; finn@lanl.gov }
\altaffiltext{4}{Theoretical Astrophysics, 
Los Alamos National Laboratory, Los Alamos, NM 87545;
colgate@lanl.gov }

\begin{abstract} 
We study the evolution of a
magnetic arcade that is anchored to an accretion
disk and is sheared  by the differential
rotation of a Keplerian disk. By including an extremely low
external plasma pressure at large distances, we
obtain a sequence of axisymmetric magnetostatic
equilibria and show that there is a fundamental
difference between field lines that are affected
by the plasma pressure and those are not (i.e.,
force-free). Force-free fields, while being twisted by the
differential rotation of the disk, expand
outward at an angle of $\sim 60^\circ$ away from
the rotation axis, consistent with the previous
studies.   These force-free field lines, however, are
enclosed by the outer field lines which originate
from small disk radii and come back to the disk
at large radii.  These outer fields experience most of  the
twist, and they are also affected most by the
external plasma pressure.
At large cylindrical radial distances, magnetic
pressure and plasma pressure are comparable so
that any further radial expansion of magnetic
fields is prevented  or slowed down greatly by
this pressure.    This hindrance to cylindrical radial 
expansion causes most of the added  twist to be
distributed on the ascending portion of the
field lines,  close to the rotation axis. Since
these field lines are  twisted most, the
increasing ratio of the toroidal $B_{\phi}$
component to the poloidal component $B_{R,z}$
eventually results in the collimation of
magnetic energy and flux around the rotation
axis. We discuss the difficulty with adding a
large number of twists within the limitations of
the magnetostatic approximation.
\end{abstract}

\section{Introduction}
\label{sec:intro}

The process of forming collimated
jets/outflows due to  disk accretion onto
central compact objects is thought to  depend on
how magnetic fields behave when they are
swirled  around by the accretion disk.
   The progress of understanding this process
has, however,  been hindered by the significant
lack of knowledge on the global magnetic field
configuration in/near the accretion disk (see
Okamoto 1999 for detailed critiques on many  MHD
models; see also Blandford 2000 for a recent
review).
     An ordered magnetic field is widely thought
to have an essential role in jet formation from a
rotating accretion disk.  Two main regimes have
been considered in theoretical models (see
Lovelace, Ustyugova
\& Koldoba 1999 for a review), the {\it
hydromagnetic regime}  where the energy and
angular momentum is carried by both the
electromagnetic field and the kinetic flux of
matter, and the {\it Poynting flux regime} where
the energy and angular momentum outflow from the
disk is carried predominantly by the
electromagnetic field. Major progress has been
made recently in the hydromagnetic regime of jet
formation, originally proposed by Blandford \&
Payne (1982).
     Dynamic MHD simulations  of the near jet
region have been carried out by several groups
(Bell 1994; Ustyugova et al. 1995, 1999; Koldoba
et al. 1995; Romanova et al. 1997, 1998; Meier
et al. 1997; Ouyed \& Pudritz 1997;
Krasnopolsky, Li, \& Blandford 1999).
   The simulation study of Ustyugova et al.
(1999)  in the hydromagnetic wind regime
indicates that the outflows are accelerated to
super-fast magnetosonic, super-escape speeds
close to their region of origin ($ \sim
80\times$ the inner radius of the disk), whereas
the collimation occurs at much larger
distances.
    Poynting flux models for the origin of  jets
were proposed by Lovelace (1976) and  Blandford
(1976) and developed further by  Lovelace, Wang,
and Sulkanen (1987), Lynden-Bell (1996), and
Colgate \& Li (1999).
    In these models the rotation of a Keplerian
accretion disk twists a poloidal field threading
the disk, and this results in outflows from the
disk which carry angular momentum (in the twist
of the field) and energy (in the Poynting flux)
away from the disk, thereby facilitating the
accretion of matter. Most recent computer
simulations using the full axisymmetric  MHD
equations have been in the hydromagnetic regime.
However, recent simulation studies have found
 jet outflows in the Poynting flux regime
(Romanova et al. 1998; Ustyugova  et al. 2000).

One class of models deals with a simplified
limit, where magnetostatic and force-free
conditions are assumed. This kind of
problem is very clearly stated in the introduction
of  Lynden-Bell \& Boily (1994, hereafter LB94;
see also  Lynden-Bell 1996): A field rooted in a
heavy conductor on $z=0$  (i.e., the rotating
disk) pervades a perfectly conducting
force-free medium in the region
$z>0$. On the disk surface ($z=0$), the field
passes upwards from the region $0 \leq R \leq
R_{\rm o}$ and returns downwards in the region
$R_{\rm o} < R \leq R_{\rm max}$. The disk is
now rotated about its axis  according to the
Keplerian motion, $\Omega(R) \propto R^{-3/2}$.
The problem is to determine the magnetostatic
field configuration when the disk has gone
through a specific number of turns.
They found that fields, instead of collimating
along the rotation axis, expand along an
angle of $\theta \approx 60^\circ$ away from the 
rotation axis. 

The behavior of twisted magnetic field lines that are
anchored in a perfectly conducting medium has originally
been considered in the solar corona context 
(see, e.g., Aly 1984, 1991, 1995; Mikic et al. 1988; 
Finn \& Chen 1990; Sturrock et al. 1995). The direct
application of those studies to accretion disks
has been fairly recent (e.g., Appl \& Camenzind 1993;
K\"onigl \& Ruden 1993; LB94; Lynden-Bell 1996;
Bardou \& Heyvaerts 1996; Goodson et al. 1999;
Uzdensky et al. 2000). 
The important role of an external plasma pressure was
first emphasized by Lynden-Bell (1996) where he
argued that field lines can not simply expand to
infinity as shown by LB94 because the required
$PdV$ work becomes too large.
He then constructed a simplified cylindrical model where a
magnetic helix/jet is bounded at large radius by
external plasma pressure. But the question left open
is whether the plasma pressure can indeed play
a confining role in a self-consistent {\em global} treatment
of the twisting and expansion of magnetic fields
driven by the disk shear rotation. 

In this study, we obtain
self-consistent global  solutions of
axisymmetric magnetostatic configurations by
strictly enforcing the Keplerian shear condition
on the flux surfaces  that are anchored in the
disk. The assumptions we use are explained in
detail in \S \ref{sec:assum}. We formulate our
problem in \S \ref{sec:eqs}, along with the
relevant equations and methods to solve them.
Results are given \S \ref{sec:result} and final
conclusions and discussions are in \S
\ref{sec:diss}.

\section{Basic Assumptions}
\label{sec:assum}

The equation of motion in the non-relativistic
ideal magnetohydrodynamics (MHD) limit is simply
\begin{equation}
\label{eq:motion}
\rho \frac{d{\bf v}}{dt} = \frac{1}{c}~{\bf J
\times B} + \rho {\bf g}  - \nabla P~,
\end{equation} where $\rho$ is the plasma
density, ${\bf v}$ the flow velocity,
${\bf J}$ the current density,
${\bf g}$ the gravitational acceleration, and
$P$ is the plasma pressure.

We restrict our attention to axisymmetric 
magnetostatic configurations.
The underlying assumption is that the speed
of the field line twisting by the disk is slow
so that the system  quickly reaches an equilibrium.
We can then treat the evolution of field
configurations  as sequences of magnetostatic
equilibria.   
We make the further assumption that the magnetic
field plays the  dominant role with plasma
flow ${\bf v}$ and gravity $\rho {\bf g}$ much
less important.  Then, the steady state equation is
\begin{equation}
\label{eq:ff1} {\bf J \times B} = c~ \nabla P~~.
\end{equation} 
The main astrophysical motivation for keeping this 
pressure term is that at sufficiently large distances 
away from the disk, the plasma  pressure will be 
comparable to the magnetic pressure, thus becoming 
dynamically important.
When $ \nabla P \rightarrow 0$,
we have the so-called force-free (FF) limit.

We now discuss the physical conditions where
these assumptions apply. There are at least two
relevant speeds in this problem which are
related to two physical aspects.
The first one is the velocity of Keplerian
disk rotation, $v_{\rm K}$,  which describes the
rate of field line footpoint movement.
The second is the relaxation speed, $v_{\rm
X}$, for magnetic fields  to reach equilibrium,
which is usually achieved by MHD waves going
back and forth in the system.
This is essentially the Alfv\'en speed $v_{\rm A}$ and
is ultimately  limited by the speed of light
$c$. Together with these speeds, there are two
relevant timescales: the disk rotation period
$T_{\rm K} = 2\pi R_{\rm min}/v_{\rm K}$, and
the relaxation time $T_{\rm X} = L_{\rm
max}/v_{\rm X}$, where $R_{\rm min}$ is the
inner disk radius, and $L_{\rm max}$ the system
dimension, respectively. In order to ensure the
steady state condition on the timescale of 
$T_{\rm K}$, one requires that 
$L_{\rm max} \ll (v_{\rm X}/v_{\rm K}) R_{\rm min}$.

In order to simulate a system that is much larger
than $R_{\rm min}$, these considerations indicate that 
in a black hole accretion disk system, field lines close to
the black holes are likely twisted too rapidly to be 
treated by the magnetostatic equations.
In other words, the dynamic
pressure from the inertial term $\rho d{\bf
v}/dt$ in equation (\ref{eq:motion}) is likely
to be very important during the expansion, thus
the conclusions from this study might not apply in this limit.
On the other hand, for field lines further
away from the black hole and/or for accretion
disks around systems like young stars, the
rotation speed is small enough
that the magnetostatic limit could still apply.

Another  relevant process is the footpoint drift  
across the disk.  Magnetic field threading the disk
tends to be advected inward with the accretion
flow, but at the same time it may diffuse
through the disk owing to a finite magnetic
diffusivity $\eta_m$ of the disk.
The outward drift of the magnetic
field in the disk occurs at speed
${\cal U}_r =(\eta_m/h) \tan(\theta)$,
where $h$ is the half-thickness of the disk,
$\tan(\theta) \equiv (B_r/B_z)_{z=0}$, and a smaller, 
second order diffusion term
($\eta_m\partial^2 B_z/\partial r^2$)
has been omitted (Lovelace {\it et al.} 1997).
For cases where the diffusivity
is of the order of the viscosity and
where the viscosity is given by the
Shakura and Sunyaev (1973) prescription
$\nu =\alpha c_s h$ (with $\alpha <1$ and $c_s$
the midplane sound speed), the diffusive
drift speed is  ${\cal U}_r \sim \alpha c_s
\tan(\theta)$.   This is larger than the
radial accretion speed $v_r \sim -\alpha c_s (h/R)$
due to viscosity alone.  But what is important
here is that the field drift and the accretion
speeds are much less than the Keplerian velocity
of the disk for $c_s \ll V_K$.  For this reason the fields can
be treated as frozen into the disk.
This point was made by Ustyugova et al.
(2000) and also discussed by Uzdensky et al. (2000).

\section{Basic Equations}
\label{sec:eqs}

We have formulated the problem discussed in
\S \ref{sec:intro} in both cylindrical $(R,\phi,z)$
and spherical $(r, \theta, \phi)$
coordinates.   Axisymmetry is assumed in both
cases.  The magnetic field can be written as:
\begin{equation} {\bf B}  = {\bf B}_p + B_\phi
\hat{\bf \phi} = \nabla \Psi \times
\nabla \phi + B_\phi \hat{\bf \phi}~,
\end{equation} where the poloidal component
${\bf B}_p$ is in the $\{ R,z \}$ or
$\{ r, \theta \}$ plane, 
$2\pi \Psi$ is the total poloidal flux through
the disk, and contours of $\Psi$ label the
poloidal field lines. The poloidal magnetic
fields are
\begin{equation}
\label{eq:brbzc} B_R = -{1 \over R}{\partial \Psi
\over \partial z}~,\quad B_z ={1 \over R}
{\partial \Psi \over \partial R}~,
\end{equation} for cylindrical coordinates and
\begin{equation}
\label{eq:brbzs} B_r = {1 \over
r^2\sin\theta}{\partial \Psi
\over \partial \theta}~,\quad B_\theta =-{1
\over r\sin\theta} {\partial \Psi \over \partial
r}~,
\end{equation} for spherical coordinates. The
current density
\begin{equation}
\frac{4\pi}{c} {\bf J} = - \Delta^\star \Psi
\nabla \phi + \nabla(r B_\phi) \times \nabla
\phi~,
\end{equation} where the operator $\Delta^\star$
is, for cylindrical  coordinates,
\begin{equation}
\Delta_c^\star \Psi = \frac{\partial^2}{\partial
R^2}-\frac{1}{R}
\frac{\partial}{\partial
R}+\frac{\partial^2}{\partial z^2}~,
\end{equation} and for spherical coordinates,
\begin{equation}
\Delta_s^\star \Psi = \frac{\partial^2}{\partial
r^2}  +\frac{1}{r^2}\frac{\partial^2}{\partial
\theta^2} -\frac{1}{r^2
\tan\theta}\frac{\partial}{\partial \theta}~.
\end{equation}

Equation (\ref{eq:ff1}) implies that
\begin{equation}
\label{eq:ppsi} {\bf B} \cdot \nabla P = 0 ~~
\Longrightarrow ~~ P = P(\Psi)~.
\end{equation} In other words,  the gas pressure is
constant along field lines.   In cylindrical
coordinates, equation (\ref{eq:ff1}) can be
written as the well-known Grad-Shafranov
equation,
\begin{equation}
\Delta_c^\star \Psi \nabla \Psi + RB_\phi
\nabla(RB_\phi) = - 4\pi R^2
\frac{dP}{d\Psi} \nabla \Psi~,
\end{equation} 
from which we define $RB_\phi =
H(\Psi)$ and we get
\begin{equation}
\label{eq:ff_c}
\Delta_c^\star \Psi + d[H^2/2]/d\Psi + 4\pi R^2
dP/d\Psi = 0~.
\end{equation} 
Similarly, in spherical
coordinates, we define $r\sin\theta B_\phi =
H(\Psi)$ and
\begin{equation}
\label{eq:ff_s1}
\Delta_s^\star \Psi + d[H^2/2]/d\Psi + 4\pi
(r\sin\theta)^2 dP/d\Psi = 0~.
\end{equation} 
In our spherical calculations, we
actually use $\xi = \ln r$ to concentrate the uniform grid 
in $\xi$ for better resolution at small $r$.  Equation
(\ref{eq:ff_s1}) then becomes
\begin{equation}
\label{eq:ff_s}
\Delta_{sl}^\star \Psi + r^2 d[H^2/2]/d\Psi + 4\pi
(r^4\sin\theta^2) dP/d\Psi = 0~,
\end{equation} where
\begin{equation}
\Delta_{sl}^\star \Psi =
\frac{\partial^2}{\partial \xi^2}-
\frac{\partial}{\partial
\xi}+\frac{\partial^2}{\partial \theta^2}
-\frac{1}{\tan\theta}\frac{\partial}{\partial
\theta}~.
\end{equation}

The quantity $H(\Psi) = RB_\phi(R,z)$ is $(2/c)$
times the total current flowing through a
circular area of radius $R$ (with normal
$\hat{\bf z}$) labeled by $\Psi(R,z)$= const.
The development of the toroidal field component
from an initial purely  poloidal field comes
from the differential rotation of the disk onto
which footpoints of the same field lines are
anchored.

\subsection{Boundary Conditions}

Since there is no direct observational
information on how magnetic fields are
distributed on the surface of an accretion disk,
we have made the following assumptions.  If the
magnetic fields on the disk is initially
generated via  an accretion disk dynamo that
makes a quadrupole or dipole-like field, magnetic fields
will emerge out of the disk at smaller radii
and  go back to the disk at large radii.
We use $R_{\rm o}$ to represent the
{\cal O}-point of the field in the disk, i.e., the radius where
$B_z = 0$ on the disk. Furthermore, in  a
fully self-consistent treatment including the
back-reaction on the disk, $R_{\rm o}$ also
marks a separation location inside which angular
momentum is lost and transmitted to the outer
part of the disk by field line tension.

\begin{figure}
\begin{center}
\epsfig{file=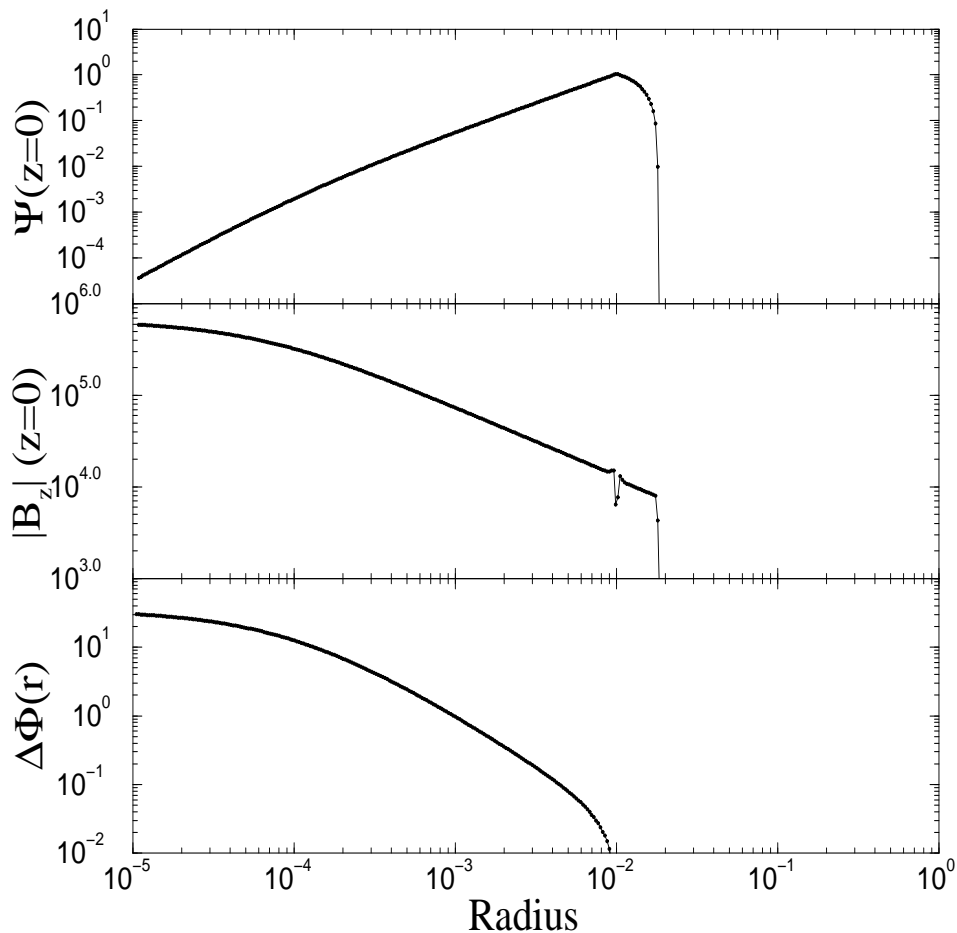,height=5in,width=5in}
\end{center}
\caption{The distribution of poloidal fields on
the disk  between $0 \leq R \leq R_{\rm out}$.
Its $\Psi$ value is shown in the top panel, which
increases from 0 to 1 at the {\cal O}-point $R_{\rm
o}=0.01$ and  decreases back to 0 at $R_{\rm
out} = 0.02$.  Its corresponding $|B_z(R,0)|$ is
shown in the middle panel, where it becomes zero
as well as reverses its sign at $R_{\rm o}$.  
The twist profile
($\propto R^{-3/2}$ is shown in the  bottom
panel. It is nearly Keplerian over large range
of radii, but is flattened for $R \rightarrow 0$
and is approaching zero near the {\cal O}-point because
the separation between the footpoints of the
arcade is becoming zero.
\label{fig:init_psibz}}
\end{figure}

We study the expansion of this cylindrical magnetic
arcade anchored on the disk. We further assume
that the overall strength of $|B_z|$  decreases
as a function of radius, except near the {\cal O}-point
where $|B_z| = 0$. This is roughly consistent
with the fact that the thermal pressure of the
disk (which anchors the fields) decreases
radially as well. Specifically, we have chosen
a computational domain that is a box with $0 < R
< R_{\rm max}$ and $0 < z < z_{\rm max}$ in the
cylindrical case and/or a sphere with 
$r_{\rm min} < r < r_{\rm max}$  in the spherical case.
The outer boundary is assumed to be perfectly
conducting ($\Psi=0$).  At $z=0$ (along the
disk), we assume that the disk is perfectly
conducting as well. An initial poloidal field
distribution is specified
$\Psi(R,0)$ [or $\Psi(r,\pi/2)$] as
\begin{equation}
\label{eq:psi_disk}
\Psi(R,0) \propto
\cases{ R^2 & for $R < R_c$ ;\cr R^\alpha & for
$R_c < R < R_{\rm o}$; \cr R_{\rm out}^\alpha -
R^\alpha & for $R_{\rm o}< R < R_{\rm out}$, \cr }
\end{equation} 
where we have joined these parts
smoothly.  Figure \ref{fig:init_psibz} shows the
distribution of $\Psi$ and $|B_z|$ on the disk
surface. The poloidal flux is distributed
between $R_{\rm min}\approx 10^{-5}$ and
$R_{\rm out}\approx 0.02$.  A core radius
$R_c\approx 10 R_{\rm min}$ is used, inside of
which the
$B_z$ component approaches a constant. The index
$\alpha$ is chosen  to be $5/4$, so that $B_z$
is  decreasing as $R^{\alpha-2}$, reversing its
sign  at the {\cal O}-point $R_{\rm o} \approx 0.01$,
and continuing decreasing  until $R_{\rm out}$.
Both $\Psi(R,0)$ and $|B_z|$ remain zero between
$R_{\rm out}$ and $R_{\rm max}=1$. These choices
for $R_{\rm o}$ and $R_{\rm out}$ ensure that
the initial magnetic field is distributed far
inside  the outer boundary $R_{\rm
max}$. We have tried many different initial
field configurations and found that our main
conclusions  do not depend on these particular
choices of parameters, as long as magnetic
fields stay away from the outer boundary $R_{\rm
max}$ or $r_{\rm max}$. The dependence on 
$\Psi(R,z=0)$ near the {\cal O}-point is very weak
since those field lines are hardly twisted at all.

\subsection{External Plasma Pressure}
\label{sec:press}

The physical picture we have in mind for
accounting for the pressure of an external plasma is that
magnetic field close to the disk is
force-free with little influence from the plasma
pressure. At large distances, however, the
magnetic pressure decreases sufficiently so that
plasma pressure becomes important.
In other words, the magnetic fields are
bounded above and  on the sides by an ambient
medium which is perfectly conducting and mostly
devoid of any magnetic flux.  This forms a
conducting, in general, ``deformed box'' inside of
which the magnetic field dominates and 
the gas pressure is very small. 

To model the external plasma pressure,
we adopt the dependence
\begin{equation}
\label{eq:press} 
P(\Psi) = P_c \exp[-(\Psi/\Psi_c)^2]~,
\end{equation} 
so that
\begin{equation} 
dP/d\Psi = -
(2P_c/\Psi_c^2)~\Psi~ \exp[-(\Psi/\Psi_c)^2]~,
\end{equation} 
where $P_c$ and $\Psi_c$ are the
input parameters. For most of the results presented here,
we have chosen
$P_c=0.1$ and $\Psi_c=10^{-3}$.  The physical
meaning of $P_c$ is that its magnitude gives an
estimate of the overall importance of
plasma pressure. Using the example given in
equation (\ref{eq:psi_disk}),  the maximum $B_z$
near $R_{\rm min}$ is $\sim 6\times 10^5$,  (see
Figure \ref{fig:init_psibz}),  so the ratio of
the maximum magnetic pressure to the maximum
gas pressure  in
the computational domain is $\sim 10^{11}$.  The
quantity $\Psi_c$ gives the fraction of poloidal
flux that is affected by the plasma pressure.

The prescribed plasma pressure (equation \ref{eq:press})
is kept fixed at all times throughout the sequence of equilibria.
This is because the prescribed pressure here is meant to
mimic a {\em constant} external pressure boundary (such as
pressure from interstellar or intergalactic medium) that 
is reacting to the ``push'' by magnetic fields. This interpretation
is precise in the limit of $\Psi_c \rightarrow 0$. We have used a 
very small but finite $\Psi_c$ for the convenience of modeling 
both the  Lorentz force and pressure gradient terms simultaneously, 
rather than having to specify the pressure gradient term through
a boundary condition, especially when the shape of this boundary 
is determined dynamically by the push from the magnetic fields
and is not known {\em a priori}.
The pressure in the flux tubes with $\Psi > \Psi_c$, however,
might change due to the flux volume expansion. If the entropy
of each flux tube is conserved, then pressure has to decrease
with expanding volume. On the other hand, if there is enough heat 
flux between the disk and the flux tubes, then the entropy of a 
flux tube is not conserved (Finn \& Chen 1990).
Either way, since the pressure in the flux tubes with
$\Psi > \Psi_c$ is already set to be exponentially small (equation
\ref{eq:press}) initially, any further decrease in pressure will
not alter our conclusions.

We make two more observations about
the plasma pressure term.  First, equation
(\ref{eq:press}) gives that, 
when $\Psi \rightarrow 0$ (i.e., at the boundary), 
the pressure $P \rightarrow P_c$ and its gradient
$dP/d\Psi \rightarrow 0$. Both features are
physically plausible. Second, it
is interesting to note that the pressure effect
enters  equations (\ref{eq:ff_c}) and
(\ref{eq:ff_s}) with a geometric factor
$R^2$ or $(r\sin\theta)^2$. This factor actually
comes from the
${\bf J \times B}$ term. It is only under the
equilibrium condition can it be ``moved'' to be
a multiplier of the pressure term.
Consequently, the pressure tends to prevent
expansion away from the rotation axis but has
relatively little constraining effect along the
rotation axis.  

\subsection{Input Keplerian Field Line Twist}

From the distribution of $\Psi(R,0)$ and the
Keplerian rotation
$\Omega(R) \propto R^{-3/2}$, we can define the
required twist on each field line $\Delta
\Phi(\Psi)$.  The twist of a given field line
going from an inner footpoint at $R_1$ to an
outer footpoint at $R_2$ is proportional to the
differential rotation of the disk.  Using a
cylindrical coordinate, for a given field line,
we have
$$ Rd\phi/B_\phi = d\ell_p/B_p~,
$$ where $d\ell_p \equiv \sqrt{dR^2+dz^2}$ is the
poloidal arc length along the field line, and
$B_p \equiv \sqrt{B_R^2+B_z^2}$. Then the total
twist of a field line is
\begin{equation}
\label{eq:path}
\Delta \Phi(\Psi) =
\int_1^2 d\ell_p ~{B_\phi \over R B_p} =H(\Psi)
\int_1^2 {d\ell_p \over R^2 B_p}~.
\end{equation} We denote the quantity $V'(\Psi)
= \int_1^2 d\ell_p/R^2 B_p$. The field line
twist after a time $t$ is
\begin{eqnarray}
\label{eq:twist}
\Delta \Phi(\Psi) &= &  \Omega_{o} ~t
\left[\left({R_{\rm o}\over R_1}\right)^{3/2} -
\left({R_{\rm o}\over R_2}\right)^{3/2}\right]~,
\nonumber \\ &=& (\Omega_{o}~t) ~{\cal F}(\Psi)
\end{eqnarray} where
$\Omega_{o}\equiv\sqrt{GM/R_{\rm o}^3}$ is the
Keplerian  angular frequency at $R_{\rm o}$ of
an object of mass $M$, and ${\cal F}$ is
dimensionless.  The rotation profile is
assumed to begin deviating from Keplerian and
approach a constant when $R \leq R_{\rm
c}$ (the bottom panel of Figure
\ref{fig:init_psibz}). There exists a region,
however, between $\Psi = 0$ and
$\Psi_{\rm min}=\Psi(R_{\rm min})$ where the
twist decreases as $\Psi$ decreases (i.e.,
getting closer to the $z$-axis and the
surrounding wall).  These field lines are not
explicitly followed in our calculations. But
since $H(\Psi) \rightarrow 0$ as
$\Psi \rightarrow 0$, the twist $\Delta
\Phi(\Psi)$ approaches zero as well
[equation (\ref{eq:path})].

\subsection{Method of Solution}

We solve equations (\ref{eq:ff_c}) and
(\ref{eq:ff_s}) following the approach given in
Finn \& Chen (1990), which employs several
levels of iterations.  For the very first step,
we will use a trial $H(\Psi)$, then we solve
equation (\ref{eq:ff_c}) or (\ref{eq:ff_s}) 
using the Successive Over-Relaxation (SOR) method
(see for example, Potter 1973)
for $\Psi^k(r,\theta) \rightarrow
\Psi^{k+1}(r,\theta)$. From
$ \Psi^{k+1}(r,\theta)$, we integrate along the
field lines to obtain
$V'(\Psi)$. We typically trace $\sim 200$ field lines 
with $3\times 10^{-5} \leq \Psi \leq 1$ for calculating
the poloidal current profile $H(\Psi)$. 
Using equation (\ref{eq:path}), we
update $H(\Psi)$ using the input $\Delta
\Phi(\Psi)$. This procedure is then repeated.
The advantage of the simple SOR solver is that
the nonlinear source  terms $d(H^2)/d\Psi$ and
$dP/d\Psi$ can be easily included in the
iterations so that  convergence can be achieved
fairly quickly (with a typical relative
residue less than $10^{-6}$)

In summary, we compute global
magnetostatic solutions of
$\Psi(r,\theta)$ using the Grad-Shafranov
equation in axisymmetry by  requiring that the
twist on each field line follows a specified
function $\Delta \Phi(\Psi)$, derived based on
the Keplerian disk  rotation.

\section{Results}
\label{sec:result}

We have made many runs with different choices of
the initial  field configuration on the disk
$\Psi(R,0)$, the twist profile
$\Delta \Phi(\Psi)$, and the pressure profile
$P(\Psi)$.  Runs are made in both cylindrical
and spherical coordinates. We present most of
our results using the following parameters: A
$\ln r-$spherical coordinate system is used, with
$r_{\rm min} = 10^{-5}, r_{\rm max} = 1$.  The
radial and $\theta-$angle grids are $367\times 51$. 
The initial field configuration is the
same as shown in Figure \ref{fig:init_psibz},
where $R_{\rm o} = 0.01$ and $R_{\rm out}=
0.02$.  The smallest $\Psi$ value of the field
lines we track is $\Psi_{\rm min} = \Psi(R_{\rm
min}) = 3\times 10^{-5}$. The plasma pressure
parameters are $P_c = 0.1$ and $\Psi_c = 0.001$.
The maximum $|B_z|$ on the disk is $\approx
6\times 10^5$,  which implies a ratio of $\sim
10^{11}$ between the maximum magnetic pressure
and the plasma pressure. (Even larger pressure
ratios can be achieved with relative ease.) To
indicate the progressively increasing twist
added to the field lines, we use the notation
time ``$t$'' [equation (\ref{eq:twist})] in
units of revolutions of the inner most flux line
($\Psi_{\rm min}$) around the $z$-axis, i.e.,
$\Delta \Phi_{\rm max} = \Delta \Phi(\Psi_{\rm
min}) = 2\pi t$.

\begin{figure}
\begin{center}
\epsfig{file=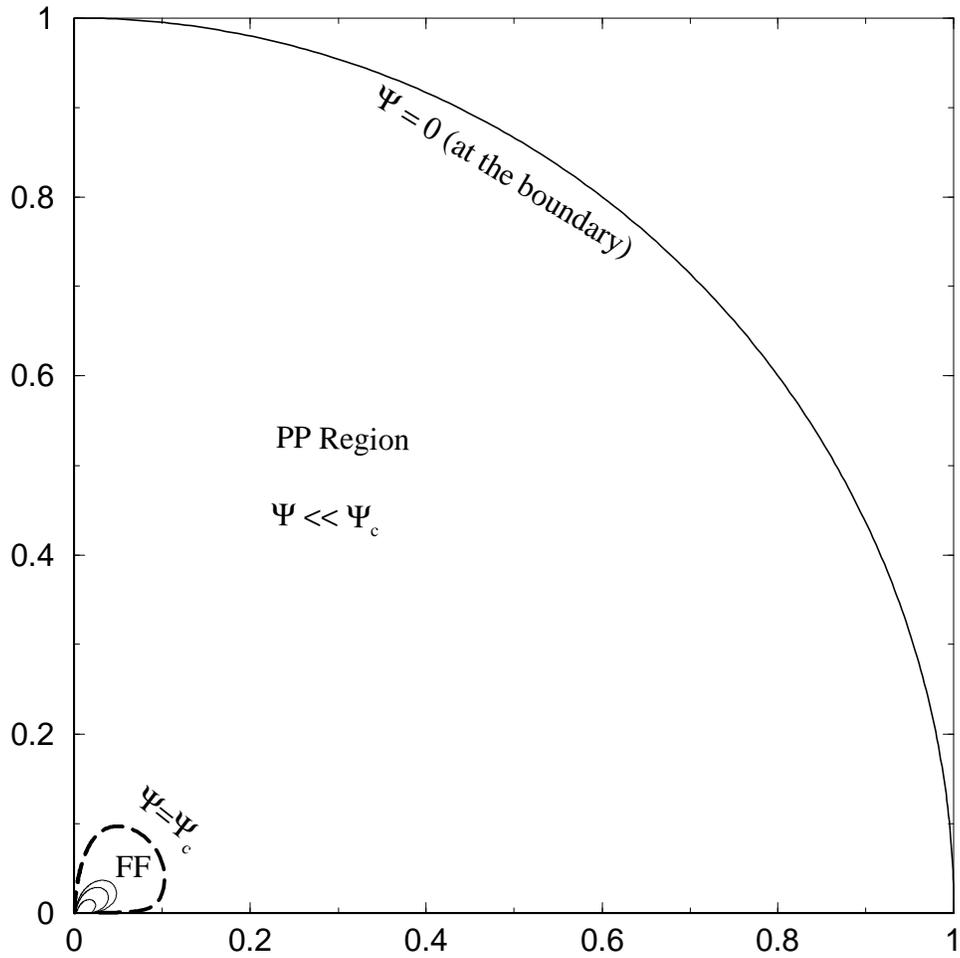,height=5in,width=5in}
\end{center}
\caption{The overall physical picture with two
different regimes, separated by $\Psi = \Psi_c$
(the thick dashed line). The region labeled
``plasma-pressure'' (PP) is dominated by the 
plasma pressure with
$\Psi_c \geq \Psi \geq 0$ (the boundary at $r =
1$). A tiny fraction (=$\Psi_c$) of the total
poloidal flux is in this region. The region
labeled ``force-free'' (FF) is dominated by the magnetic
pressure  with $\Psi_c \leq
\Psi \leq 1$ (the {\cal O}-point). Most of the poloidal
flux ($1-\Psi_c$) is contained in this region.
The key question we are addressing is the
response of the $\Psi=\Psi_c$ boundary to the twisting
of field lines  by the Keplerian disk.
\label{fig:model}}
\end{figure}

Figure \ref{fig:model} describes our overall
physical picture with two different physical
regimes. For $0 \leq \Psi \leq \Psi_c$, the
plasma pressure is dominant. This region is
labeled as ``plasma-pressure'' (PP). 
For $\Psi_c \leq \Psi \leq
1$, the magnetic pressure dominates, the region
is force-free (labeled as ``FF''). So, we have
effectively made two ideal MHD fluids, one (PP)
has  a plasma $\beta \rightarrow \infty$ and the
other (FF) has $\beta = 0$. 
The key question we are addressing is:
``how does the shape of
$\Psi=\Psi_c$ boundary change while field lines
are being twisted by the Keplerian disk
rotation?'' In other words,
the boundary between the PP and FF
regions  evolves according to both the expansion and
``pushing'' from the FF region and the
``hindrance'' of the plasma pressure.

Although some details may differ, we generally
find that  magnetic fields evolve in the
following manner: (1) in the FF region, field lines
expand  primarily towards the large radius along
an angle (from the
$z$-axis) of $\theta \sim 45^\circ - 60^\circ$;
(2) in the PP region, field lines expand
predominantly vertically (along the
$z$-axis) with some slight radial expansion; (3)
more twist causes further expansion along the
$z$-axis but  subject to a break-down of the
magnetostatic, equilibrium  assumption.  We now
discuss these features in detail.

\subsection{Equilibrium Sequence with Increasing
Twist}

Figure \ref{fig:seq} shows the evolution of
magnetic fields in the poloidal plane  as twist
is added, with $t = 0, 1, 2$ and $4$ turns,
respectively. The field lines shown are evenly
spaced in
$\log_{10}(\Psi)$, with the outer-most line
having $\Psi \approx 10^{-4}$ and the inner-most
line having $\Psi \approx 0.5$. Here, we use the
term ``outer'' to refer to field lines in the PP
region, which originate from the  smallest radii
and return back to disk at the largest radii.
The term ``inner'' refers to field lines around
the {\cal O}-point in the FF region.   Consequently,
because the field lines are twisted according to
the Keplerian rotation (largest rotation at the
smallest radius),  the outer field lines show
substantial ``movement'' due to the added
twist, whereas field lines around the {\cal O}-point
show little change.

The corresponding results for the poloidal
current $H(\Psi)$ and the resultant twist
$\Delta \Phi$ are plotted in Figure
\ref{fig:htp}. It is interesting to note that
$H(\Psi)$ can be approximated by two power-laws
of $\Psi$, where $H(\Psi) \propto \Psi$ when
$\Psi \leq \Psi_c$ and  flattens somewhat for
$\Psi > \Psi_c$.  The twist shown in the lower
panel is derived by  integrating along each
field line. They indeed follow closely the
input profile due to Keplerian rotation.

\subsection{Role of External
Plasma Pressure in Collimation}

\begin{figure}
\begin{center}
\epsfig{file=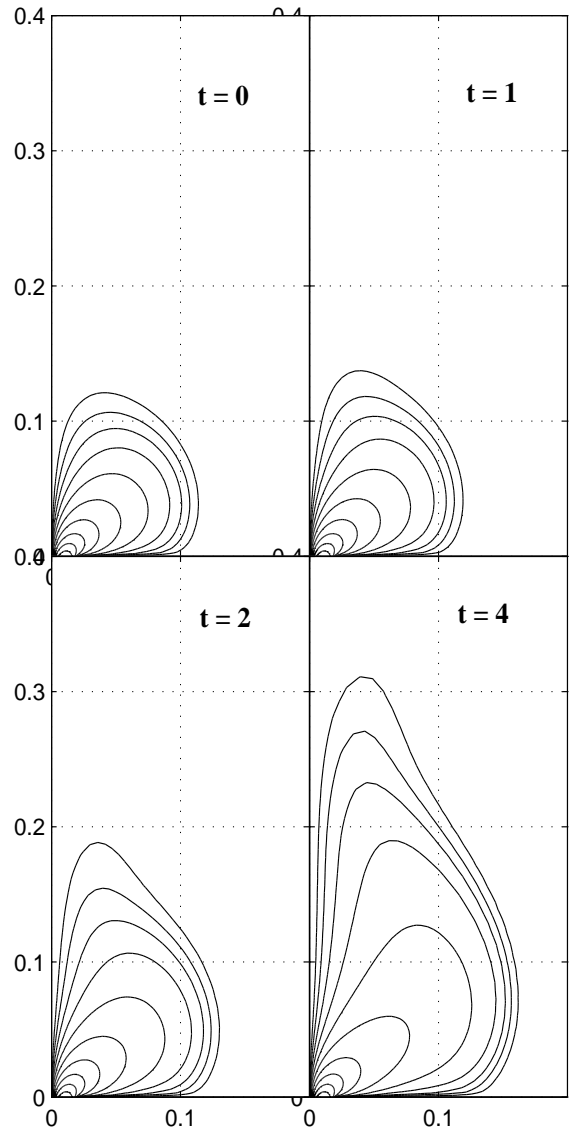,height=6in,width=3in}
\end{center}
\caption{The ``evolution'' of the poloidal field
lines  with increasing twist as solutions of
equation (\ref{eq:ff_s})  in a $\ln r-$spherical
coordinate for $t = 0, 1, 2$ and $4$,
respectively. (We present the results in a
smaller $R-z$  plane for clarity.) The contours
are displayed evenly in logarithmically spaced
intervals ($10^{-4} \leq \Psi \leq 1$). The
outer field lines  have expanded more strongly
along the $z-$axis (from $z \approx 0.1$ to
$0.3$) than in the radial direction(from $R
\approx 0.1$ to $0.15$).
\label{fig:seq}}
\end{figure}

\begin{figure}
\begin{center}
\epsfig{file=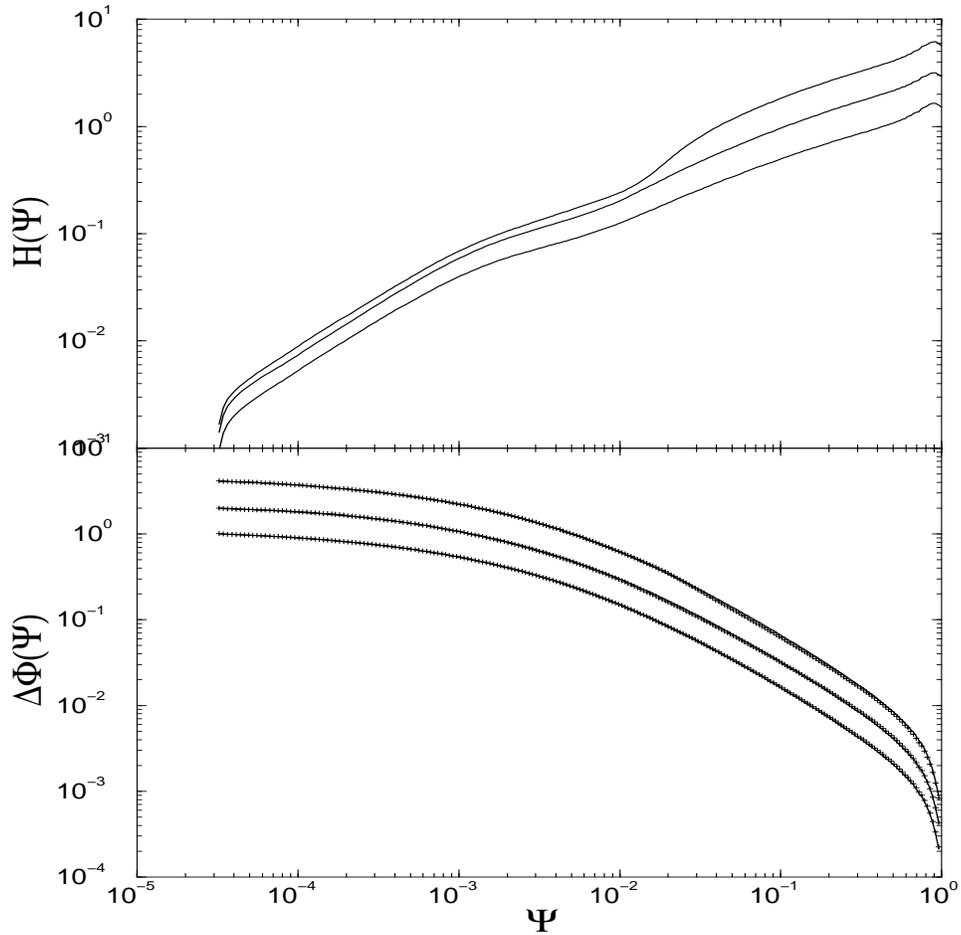,height=5in,width=5in}
\end{center}
\caption{The poloidal current $H(\Psi)$ (upper
panel) and  the twist $\Delta \Phi(\Psi)$
(lower panel) as a function of $\Psi$ for
$t=1,2,4$, (bottom to top curves),
respectively, for the run shown in Figure \ref{fig:seq}.
The function $H(\Psi)$ is
$\propto \Psi$ for small $\Psi$ then  flattens
as $\Psi$ increases. The twist agrees with the
input Keplerian profile nearly perfectly with
indistinguishable differences.
\label{fig:htp}}
\end{figure}

\begin{figure}
\begin{center}
\epsfig{file=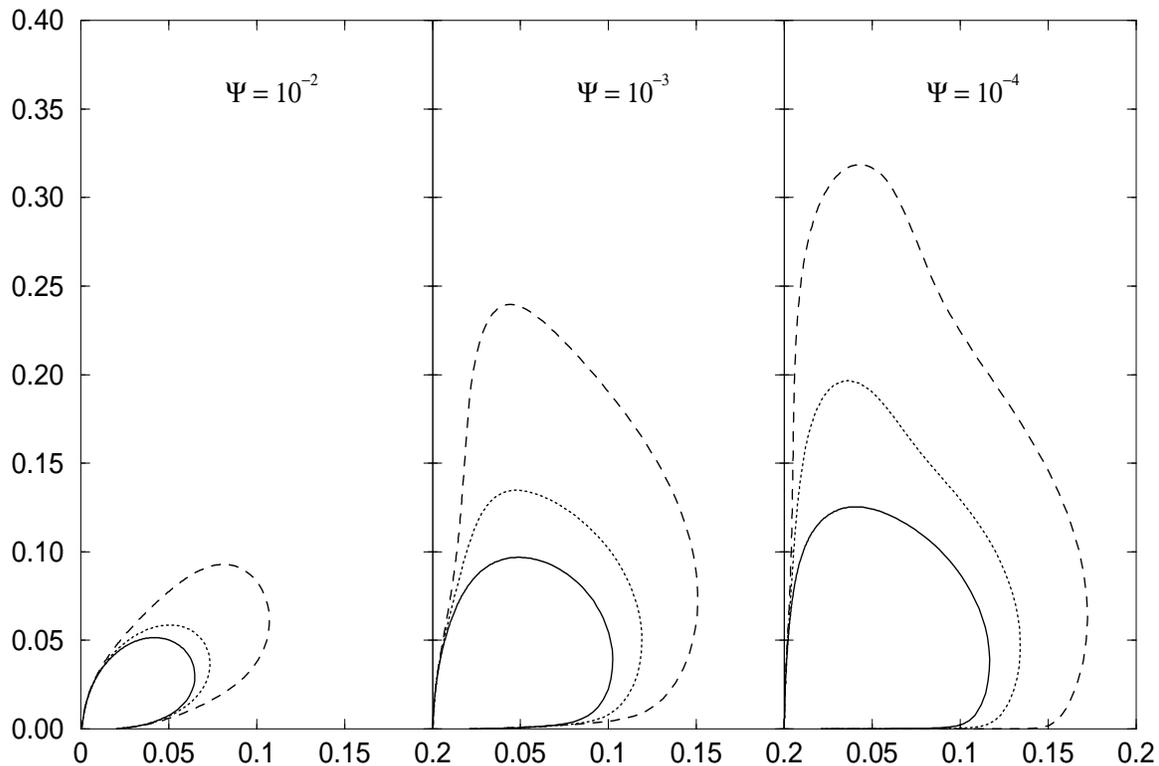,height=6in,width=4in,angle=-90}
\end{center}
\caption{The ``movement'' of three particular
poloidal field lines,
$\Psi = 10^{-2}$ (left), $10^{-3}$ (middle), and
$10^{-4}$ (right), as their twist increases from
$t=0$ (solid), $2$ (dotted) to $4$  (dashed), using
the results from Figure \ref{fig:seq}.
The plasma pressure is important for $\Psi <
\Psi_c = 10^{-3}$.   For field lines with $\Psi
> \Psi_c$, they are force-free and  expand along
an angle $\theta \sim 60^{\circ}$ radially (the
left panel). For field lines with $\Psi \leq
\Psi_c$, their radial expansion is much
``slower'' than their vertical expansion. This
results in the collimation around the $z-$axis.
Again, these solutions were obtained in a $\ln
r-$spherical coordinate with $r_{\rm max} = 1$
but we present the results  in a smaller $R-z$
plane for clarity.
\label{fig:expan}}
\end{figure}

The striking feature of the sequence of equilibria in
Figure \ref{fig:seq} is the way the outer  field
lines expand. To best understand this behavior,
we separate the field lines into two groups
according to the ratio of $\Psi/\Psi_c$. These
two groups have fundamentally different
expansion behavior,  which is illustrated in
Figure \ref{fig:expan}, where the expansion of
three different field lines ($\Psi = 10^{-2},
10^{-3}, 10^{-4}$) is shown for $t=0$ (solid),
$t=2$ (dotted), and $t=4$ (dashed),
respectively.

Field line expansion in the FF region has been
studied in detail in many previous studies (see Introduction). 
The Lorentz force is the only force available (i.e.,
the magnetic pressure gradient is balanced by the
tension force). As found for example by LB94, the
field lines  expand to large radii along an
angle of $\theta \approx 60^\circ$ from the $z-$axis. 
This is consistent with the left panel of Figure
\ref{fig:expan} (also the inner field lines in
Figure \ref{fig:seq}).

The field line expansion in the PP region, however,
is fundamentally different from the  FF regime.
This is because the magnetic pressure becomes
very small at large radii and eventually at
some distance it becomes comparable to the
surrounding plasma pressure. At this location, 
further expansion of  the magnetic fields 
is greatly hindered by the plasma pressure. 
At the same time, increasing the 
twist acts to increase the ratio of $B_\phi/B_R$.
Consequently, these field lines are
increasingly  ``pinched'' around the $z$-axis,
and they eventually expand along the $z$-axis in
a ``collimated'' fashion, as shown by the middle
and right panels of Figure \ref{fig:expan} and
the late time in Figure \ref{fig:seq}.

The middle panel of Figure \ref{fig:expan} is
particularly important. It shows how the
critical boundary at $\Psi = \Psi_c$ (see Figure
\ref{fig:model}) evolves with increasing twist.
Note that its initial shape is quasi-spherical,
but a clear collimation along the $z-$axis has
developed by $t=4$.

A small amount of poloidal flux (with
$\Psi < \Psi_c$) is in the PP region and is
twisted as well. One
can ask whether these outer field lines have
contributed importantly to forming the new shape
of  the  $\Psi=\Psi_c$ boundary.  To answer this
question, we have  performed runs with a very
small $\Delta \Phi(\Psi)$ for
$\Psi < \Psi_c$, instead of the near constant
$\Delta \Phi(\Psi)$ presented in Figure
\ref{fig:htp}. We find that as long as
$\Delta\Phi(\Psi_c)$ remains the same, its shape
does not depend on the twist profile for $\Psi <
\Psi_c$. This means that the small amount of
magnetic flux beyond the $\Psi=\Psi_c$  boundary
does not change our results noticeably.

We have also performed runs with various magnitudes
of the plasma pressure. For the purely force-free case,
we confirm the results by LB94 that the
field lines expand to large radii along an
angle of $\theta \approx 60^\circ$ from the $z-$axis
and only a small total twist (less than half a turn)
can be added before the fields are packed against the
outer boundary. With a finite pressure, however,
the collimation effect is always observed as long as the
simulation region is large enough or the pressure is
large enough so that magnetic fields
are not directly packed against the outer boundary.

Thus, the external plasma pressure, no matter
how small it may be,  plays a fundamental role
in causing the field lines to collimate along
the rotation axis. It essentially stops (or
greatly slows down) the radial expansion of the
field lines, and in the meantime allows the
build-up of the toroidal component. This
interpretation is also consistent with the fact
that field lines are closely packed at large
radius (stopped by plasma pressure) but are
loosely spaced along the $z$-axis (see the
``$t=4$'' panel in Figure \ref{fig:seq}).

The results shown in
Figures \ref{fig:seq} and \ref{fig:expan}
indicate that the solution suggested by
Lynden-Bell (1996) is  indeed possible (at least
in the early expansion stage of the  helix
formation), even though details are different.
Here, we have shown global
self-consistent solutions where the field lines are 
twisted 
according to the disk Keplerian rotation.

\subsection{Distribution of Twist and Magnetic Energy}

%Figure \ref{fig:fldl} gives a 3-D view of a
%field line ($\Psi_1 \approx 3\times 10^{-5}$)
%expanding while being twisted.  

\begin{figure}
\begin{center}
\epsfig{file=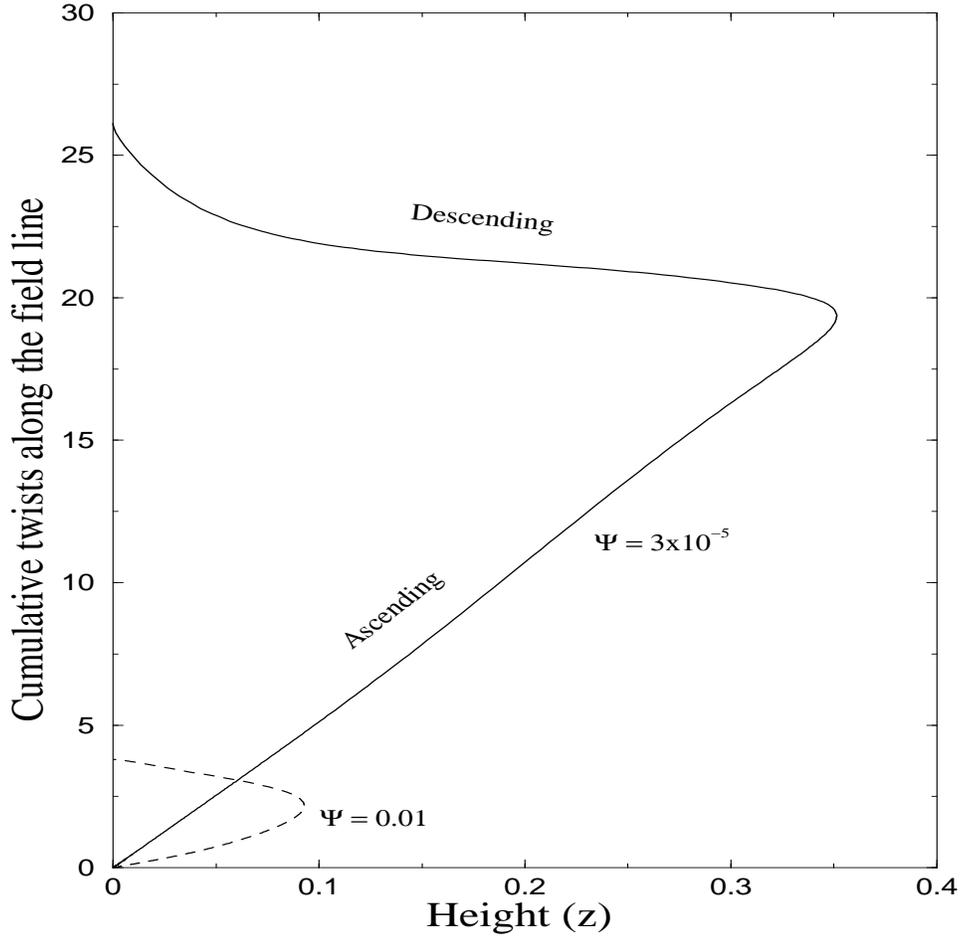,height=5in,width=5in}
\end{center}
\caption{The distribution of cumulative twist
along two field lines as a function of height
$z$. The slope of the curves $|d\phi/dz|$ gives
the  pitch of the field line (twist per unit
height).  For $\Psi=10^{-2}$ (dashed), its pitch
is equally distributed between ascending and descending portions
since it is force-free.  For $\Psi =
3\times 10^{-5}$ (solid), its twist distribution
is not uniform, with most of its total twist
being  distributed on the ascending portion of
the field line.
\label{fig:twist}}
\end{figure}

Figure \ref{fig:twist} shows how its twist ($\Delta
\phi$)  is distributed along this field line as
a function of $z$ at $t=4$. We have plotted
another field line with a larger $\Psi_2
=10^{-2}$ at $t=4$ as well. 
There is an
important difference between these  two field
lines in the distribution of the twist $\Delta \Phi(\Psi)$.
For $\Psi_1$,  most of its twist ($\geq 75\%$) is
distributed on the ascending portion  of the
field line and near the $z$-axis; the rest is
distributed while the field line spirals down
back to the disk. 
For $\Psi_2$, however, the
twist is distributed uniformly between ascending
and descending portions. To
understand this difference, we can write
$$
\left |\frac{d\phi}{dz}\right| =
\left|\frac{RB_\phi}{R^2 B_z}\right| =
\left |\frac{H(\Psi)}{R\partial \Psi/\partial
R}\right|~.
$$  
 Note that $H(\Psi)$ is constant along a
field line. Imagine a small flux tube ($\Psi
\rightarrow \Psi+d\Psi$) originating from the
inner region of the disk. Since the field line remains 
at small $R$ for the ascending section in the presence 
of plasma pressure, $B_z R^2 \sim 2 \Psi $.  This is the 
reason why $d\phi/dz$ following a flux tube is roughly 
constant for the ascending part of the flux tube.  For
the $\Psi_1$ flux tube, however, its expansion at large 
radius (the descending portion) is strongly constrained 
by the plasma pressure, so that the cross section area 
for the descending flux tube is much smaller than it would 
have been without the plasma pressure. Consequently, 
its $B_z$ component on the descending portion
of the $\Psi=\Psi_1$ surface varies more slowly than $R^{-2}$, 
so that the rate $|d\phi/dz|$ is smaller. Overall, the twist is then
non-uniformly distributed on a flux line. (See Ch .9 of Parker
1979). Thus, the presence of plasma pressure leads to a strong
collimation and a concentration of the twist to the 
collimated region.

The total injected toroidal flux ${\cal
F}_{\phi}$ by the  disk rotation into the system
can be evaluated as
\begin{equation} 
{\cal F}_{\phi}(t) = \int \int
B_\phi dR dz =
\int_0^1 d\Psi \Delta \Phi(\Psi,t)~.
\end{equation} 
We find that the ratio of ${\cal
F}_{\phi}$ to the initial total  poloidal flux,
${\cal F}_{z} = \int_0^{R_{\rm o}}dR 2\pi R B_z
= 2\pi$, increases linearly with time as
expected by equation (\ref{eq:twist}), reaching
a maximum of $\sim 4\%$ when $t=4$. So, the
total injected toroidal flux is still a small
fraction of the total initial poloidal flux.
This information is useful for non-axisymmetric
stability considerations for future studies.

%\subsection{Distribution of Magnetic Energy}

Figure \ref{fig:b2} shows the spatial distribution of
magnetic energy density and how it evolves with added
twist. The quantity $\log_{10}(B^2/P_c+1)$ 
(i.e., a scaled magnetic energy density)
is shown for $t=0$ (top left) and $4$ (top right), 
where $B^2 = B_p^2 + B_{\phi}^2$ and $P_c=0.1$. 
It is clear that  essentially all
the magnetic energy is enclosed by the plasma
pressure.  The expansion
results in a clear collimation of magnetic
energy along the $z-$axis. The two bottom panels
show, at $t=4$, the poloidal $\log_{10}(B_p^2/P_c+1)$ 
(lower left) and the toroidal
$\log_{10}(B_{\phi}^2/P_c+1)$ (lower right) components,
respectively. Note the dramatic
increase of magnetic energy (mostly poloidal
component) along the $z-$axis.   

%Figure \ref{fig:bcomps} shows the radial distribution
%of the three field components at a fixed height
%$z=0.1$ for $t=0$ (dashed) and $4$ (solid),
%respectively. Combining Figures \ref{fig:b2} and
%\ref{fig:bcomps}, we can see that, at small
%cylindrical radius,
%$B_z$ is nearly constant and is the largest
%component and it has drastically increased with
%the added twist. The $B_\phi$ component becomes
%larger than $B_p$ when sufficiently away from
%the $z-$axis. The combination of the plasma
%pressure at large radii with the newly
%generated  toroidal component at intermediate
%radii acts to ``concentrate'' the poloidal
%component along the $z-$axis, resulting in the
%collimation. All this is consistent with the results
%in the previous sections.

\subsection{Limitation on Increase of Twist}

We find that it is not possible to increase the twist
beyond what is shown in Figure \ref{fig:seq}. When more
twist is added, the system evolves from a
configuration with  all field lines being tied
to the disk to a new topology with  some
poloidal field lines close on themselves instead
of connecting to the disk (i.e., forming an
isolated ``island'' of poloidal fluxes  or
``plasmoid''). At this point, our method used to
solve equation  (\ref{eq:ff_s}) becomes
unstable. The plasmoid formation  is 
associated with the field lines that are in the
FF regime ($\Psi > \Psi_c$). 
The transition to
the plasmoid formation is  sudden,  giving
a sense of eruptive behavior of the solution.
However, as discussed below, this
numerical behavior is associated with instability of 
the method but not necessarily related to instability 
of the MHD equations.

Such eruptive behavior has been the subject of
intense research in the solar flare community
(Aly 1984, 1991, 1995; Sturrock et al. 1995).
The formation of a plasmoid from a single arcade (Inhester et al.
1992) is a loss-of-equilibrium bifurcation related to tearing
instability of the current sheet which forms at the center of the
arcade when it is sheared strongly (Finn et al. 1993).  In addition,
it has been shown that plasmoid formation can occur directly as a
consequence of linear instability when multiple arcades exist
(Mikic et al. 1988; Biskamp et a., 1989; Finn et al. 1992).

This behavior of the solutions with increasing twist
can be traced to a  mathematical nature of the
elliptical equation we are solving (e.g.,
equation (\ref{eq:ff_c}) and (\ref{eq:ff_s})).
We can write this equation
in a  pseudo-time-dependent form, e.g.,
\begin{equation}
\label{eq:dpdt}
\frac{\partial \Psi}{ \partial t} = \Delta^\star \Psi +
\frac{1}{2} \frac{d H^2}{d\Psi}  - A R^2 \Psi
\exp[-(\Psi/\Psi_c)^2]~,
\end{equation}  
where a steady state is achieved
progressively when $\partial \Psi/\partial t
\rightarrow 0$. Here, $A=8\pi P_c/\Psi_c^2$ is a
positive constant.  The computational domain can
then be  divided into three parts depending on
the ratio of $\Psi/\Psi_c$. Region I: large $R$.
The pressure term ``$-R^2 |dP/d\Psi|$''
dominates. Since it is negative, it drives
$\Psi$ to $0$ exponentially  for small $\Psi$
(i.e., $d\Psi/dt \propto - C_1 \Psi$, with $C_1
> 0$).  The solution in this region is quite
stable. Region II: small $R$ and small $z$
(close to the disk), i.e., the FF region with a
negligible pressure term.  Region III: small $R$
but large $z$ (along the rotation axis),  where
all three terms contribute.  The eruptive
behavior we found could occur in both regions II
and III, whenever the poloidal current term
$dH^2/d\Psi$ exceeds a certain critical value.
With a large current, magnetic fields tend to
expand enormously and fill up the whole
computational domain.

An important question is whether this eruptive behavior actually  
occurs in a system described with the full set of dynamical equations.
Recent axisymmetric simulations using the full set of MHD equations 
by Ustyugova et al. (2000) and Goodson et al. (1999) indicate that 
field lines can reconnect and become open. 
Clearly, the rate of reconnection is enhanced by the artificially large  
resistivity in such codes; the exact role of resistivity in allowing 
reconnection when it might not otherwise occur is not completely clear.
In Ustyugova et al. (2000), 
two regimes have been found after $\sim$ tens 
of rotation periods of the inner disk: a hydromagnetic outflow from 
the outer part of the disk, and a Poynting outflow, which has 
negligible mass flux but is dominated by the electromagnetic field 
along the rotation axis.

\section{Discussions and Conclusions}
\label{sec:diss}

We have shown that a static plasma pressure  is
fundamental in shaping the overall magnetic
equilibrium, despite the fact that the plasma
pressure is exceedingly small (the maximum
magnetic pressure over the plasma pressure is
$10^{11}$ in this study). This is essentially
different from the pure force-free models
(e.g., LB94; see however Lynden-Bell 1996). This
difference  comes from the
fact that the field lines that are being twisted
the most expand the furthest so that they are most
affected by the plasma  pressure.
 This is  opposite to the behavior in the solar
flare models where most of the shear is
concentrated around the {\cal O}-point. The least
amount of shear is applied near the {\cal O}-point in
the  accretion disk case. 
In regions where the
plasma pressure is negligible, the physics of
expansion is dominated by the force-free
condition (${\bf J\times B} = 0$). Their
expansion is predominantly towards larger radii
along an angle of $\theta \sim 60^{\circ}$ from the $z-$axis
(LB94). But for field lines affected by the
plasma pressure, their radial expansion (away from the $z-$axis)
is slowed or stopped by this pressure. The
twisting of those field lines causes them to
expand much more strongly along the $z-$axis due
to the build-up of the toroidal field component $B_{\phi}$.
The build-up of $B_{\phi}$ is
non-uniform along those field lines, with more
twist distributed on the ascending portion of
the field lines. This is, again, a direct result
of external plasma pressure.

We have obtained magnetostatic equilibria
up to a maximum of $4$ turns of the inner most field
line. For larger twist the solutions exhibit
a change  of topology and our method
breaks down. The larger values of twist should be treatable
by including the inertia term $\rho d {\bf v}/dt$
in the full set of dynamic MHD equation.
  
An important question is the
stability/instability of these
axisymmetric equilibria.
Instabilities in 3D
involving flux conversion between toroidal
and poloidal components is probably
unavoidable and this effect will be
important astrophysically:
this is because the initial poloidal flux on the
disk is probably too small to  be responsible
for the observed total magnetic flux in many
astrophysical  systems (Colgate \& Li 2000).
   The flux multiplication
by the disk rotation/twisting, and
subsequent flux conversion, could however
generate enough flux.

\acknowledgements{We acknowledge many useful
discussions with Drs. R. Kulsrud, M. Romanova, R. Rosner, 
and P. Sturrock.  We thank an anonymous referee
whose insightful comments have helped both in the clarification
of the basic assumptions and the overall presentation of the 
paper.  Part of this work
was presented at an Aspen summer workshop on
Astrophysical Dynamos in 2000.
  This research was
performed under the auspices of the U.S.
Department of Energy, and was supported in part
by an IGPP/Los Alamos grant and the Laboratory
Directed Research and Development Program  at
Los Alamos.
RVEL was supported in part by NASA grants
NAG5-9047 and NAG59735 and by NSF grant AST-9986936.
}

\begin{figure}
\begin{center}
\end{center}
\caption{The magnetic pressure distribution at
$t=0$ (top left) and $t=4$ (top right). The
quantity plotted is
$\log_{10}(B^2/P_c+1)$, where $B^2 = B_p^2 +
B_\phi^2$  is the total magnetic pressure,
$B_p^2 = B_R^2 + B_z^2$ is the poloidal
component and $P_c = 0.1$. The lower left and
right panels are for $\log_{10}(B_p^2/P_c+1)$
and $\log_{10}(B_\phi^2/P_c+1)$, respectively,
at $t=4$.  The dark blue color represents the
region with $\log_{10}(B^2/P_c+1) \approx 0$, 
i.e., $B^2 \ll P_c$. The added twist causes the
fields to expand along the $z-$axis, with
increased poloidal and toroidal magnetic
pressures. Again, these solutions were obtained
in a $\ln r-$spherical coordinate with $r_{\rm
max} = 1$ but we present the results  in a
smaller $R-z$ plane for clarity.
\label{fig:b2}}
\end{figure}

\end{document}